\documentclass[fleqn,11pt,twoside]{article}
\usepackage{espcrc1}

\usepackage{graphicx}
\usepackage[figuresright]{rotating}

\newcommand{\AmS}{{\protect\the\textfont2
  A\kern-.1667em\lower.5ex\hbox{M}\kern-.125emS}}
\newcommand{\sqrtsNN}{\mbox{$\sqrt{\mathrm{s}_{_{\mathrm{NN}}}}$}}

\title{Pion Interferometry in AuAu Collisions at {\sqrtsNN}=200 GeV}

\author{M. L\'{o}pez Noriega\address{The Ohio State University, 174 W 18th Ave., Columbus, Ohio 43210, USA}
for the STAR Collaboration
}

\begin{document}

\maketitle

\begin{abstract}
We present preliminary results from a two-pion intensity
interferometry analysis from Au+Au collisions at {\sqrtsNN}=200
GeV measured in the STAR detector at RHIC. The dependence of the
apparent pion source on multiplicity and transverse momentum are
discussed and compared with preliminary results from d+Au and p+p
collisions at the same beam energy.
\end{abstract}

\section{Introduction}

Two particle intensity interferometry (HBT) is a useful tool to
study the space-time geometry of the particle emitting source in
heavy ion collisions \cite{BAUER92,HEINZ99}. It also contains
dynamical information that can be explored by studying the
transverse momentum dependence of the apparent source size
\cite{PRATT84,MAKHL88}. Extracted parameters in HBT analysis from
Au+Au collisions at {\sqrtsNN}= 130 GeV at the Relativistic Heavy
Ion Collider (RHIC) did not agree with predictions of hydrodynamic
models that gave an almost perfect description of the
momentum-space structure of the emitting source and elliptic flow
\cite{HEINZ02}. This ``HBT puzzle" could originate from the fact
that the extracted timescales (\textit{emission duration} from the
$R_{o}/R_{s}$ ratio and \textit{evolution duration} from the
$m_{T}$ dependence of $R_{l}$) are smaller than those predicted by
the hydrodynamical model \cite{HEINZ02}.

In this paper we present two-pion correlation systematics as a
function of the transverse total mass ($m_{T} = \sqrt{k_{T}^2 +
m^{2}}$, $\textbf{k}_{T} = \frac{1}{2}(\textbf{p}_{1} +
\textbf{p}_{2})_{T}$) and multiplicity in Au+Au collisions at
{\sqrtsNN}= 200 GeV produced by RHIC at Brookhaven National
Laboratory and measured by the STAR detector. We also discuss a
fitting procedure in which the strength of the final state Coulomb
interaction between the two charged pions is taking into account
in the fit itself.

\section{Experimental Details}

Experimentally, two-particle correlations are studied by
constructing the correlation function
$C_{2}$(\textbf{q})~=~A(\textbf{q})/B(\textbf{q}). Here
A(\textbf{q}) is the measured distribution of the momentum
difference \textbf{q}~=~$\textbf{p}_{1}$~-~$\textbf{p}_{2}$ for
pairs of particles from the same event, and B(\textbf{q}) is the
corresponding distribution for pairs of particles from different
events. For this analysis we selected events with a collision
vertex position within $\pm$25~cm measured from the center of the
4~m long STAR Time Projection Chamber (TPC), and we mixed events
only if their longitudinal primary vertex positions were no
farther apart than 5~cm. We divided our sample into six centrality
bins, where the centrality was characterized according to the
measured multiplicity of charged particles at midrapidity. The six
centrality bins correspond to 0-5$\%$ (most central), 5-10$\%$,
10-20$\%$, 20-30$\%$, 30-50$\%$ and 50-80$\%$ (most peripheral) of
the total hadronic cross section. Charged pions were identified by
correlating their specific ionization in the gas of the TPC with
their measured momentum \cite{ACKER03}.

The effects of track-splitting (reconstruction of a single track
as two tracks) and track-merging (two tracks with similar momenta
reconstructed as a single track) were eliminated as described in
\cite{ADLER01}. A new procedure to take into account final state
Coulomb interaction is described in the next section.

The effect of the single-particle momentum resolution ($\delta$p/p
$\sim$ 1$\%$ for pions) induces systematic underestimation of the
HBT parameters. Using an iterative procedure \cite{ADLER01}, we
corrected our correlation functions for finite resolution effects.
The correction due to the uncertainty on the removal of the
artificial reduction of the HBT parameters associated with the
anti-merging cut has been calculated in \cite{ADAMS03} and is
included as systematic error.

\section{Fitting procedure}

The three-dimensional correlation functions were generated. The
relative momentum was measured in the longitudinal co-moving
system (LCMS) frame, and decomposed according to the Pratt-Bertsch
\cite{PRATT90,BERTS89} ``out-side-long" parametrization. There is
a Coulomb interaction between emitted particles that needs to be
taken into account in order to isolate the Bose-Einstein
interaction. This Coulomb interaction, repulsive for like-sign
particles, causes a reduction on the number of real pairs at low
q, reducing the correlation function. In our previous analysis
\cite{ADLER01,MLN02} as well as in previous experiments, this was
corrected by applying a pair Coulomb correction to each pair in
the background \cite{PRATT90} corresponding to a spherical
Gaussian source of a given radius; we call this \textit{standard}
procedure. The correlation function was then fit with the
functional form:
C($q_{o}$,~$q_{s}$,~$q_{l}$)~=~$\frac{A(\textbf{q})}{B(\textbf{q})\times
K_{coul}(q_{inv})}$~=~1~+
$\lambda$~$\exp$(-$R_{o}^{2}$~$q_{o}^{2}$~-~$R_{s}^{2}$~$q_{s}^{2}$~-~$R_{l}^{2}$~$q_{l}^{2}$),
where $K_{coul}(q_{inv})$ is the square of the Coulomb
wave-function. However, this procedure overcorrects the
correlation function since all background pairs are corrected,
including those that are not formed by primary pions.

We have implemented a new procedure, first suggested by Bowler
\cite{BOWL91} and Sinyukov \cite{SINY98} and recently used by the
CERES collaboration \cite{ADAMO03}, in which the strength of the
Coulomb interaction is taken into account in the fit itself and
only pairs with Bose-Einstein interaction are considered to
Coulomb interact; we call this \textit{Bowler-Sinyukov} procedure.
The fit in this case is:
C($q_{o}$,~$q_{s}$,~$q_{l}$)~=~$\frac{A(\textbf{q})}{B(\textbf{q})}$~=~(1-$\lambda$)~+
$\lambda~K_{coul}(q_{inv})(1~+~\exp$(-$R_{o}^{2}$~$q_{o}^{2}$~-~$R_{s}^{2}$~$q_{s}^{2}$~-~$R_{l}^{2}$~$q_{l}^{2}$)),
where $K_{coul}(q_{inv})$ is the same as above.

\begin{figure}
\begin{minipage}[t]{0.45\textwidth}
\includegraphics[width=\textwidth,height=0.65\textwidth]{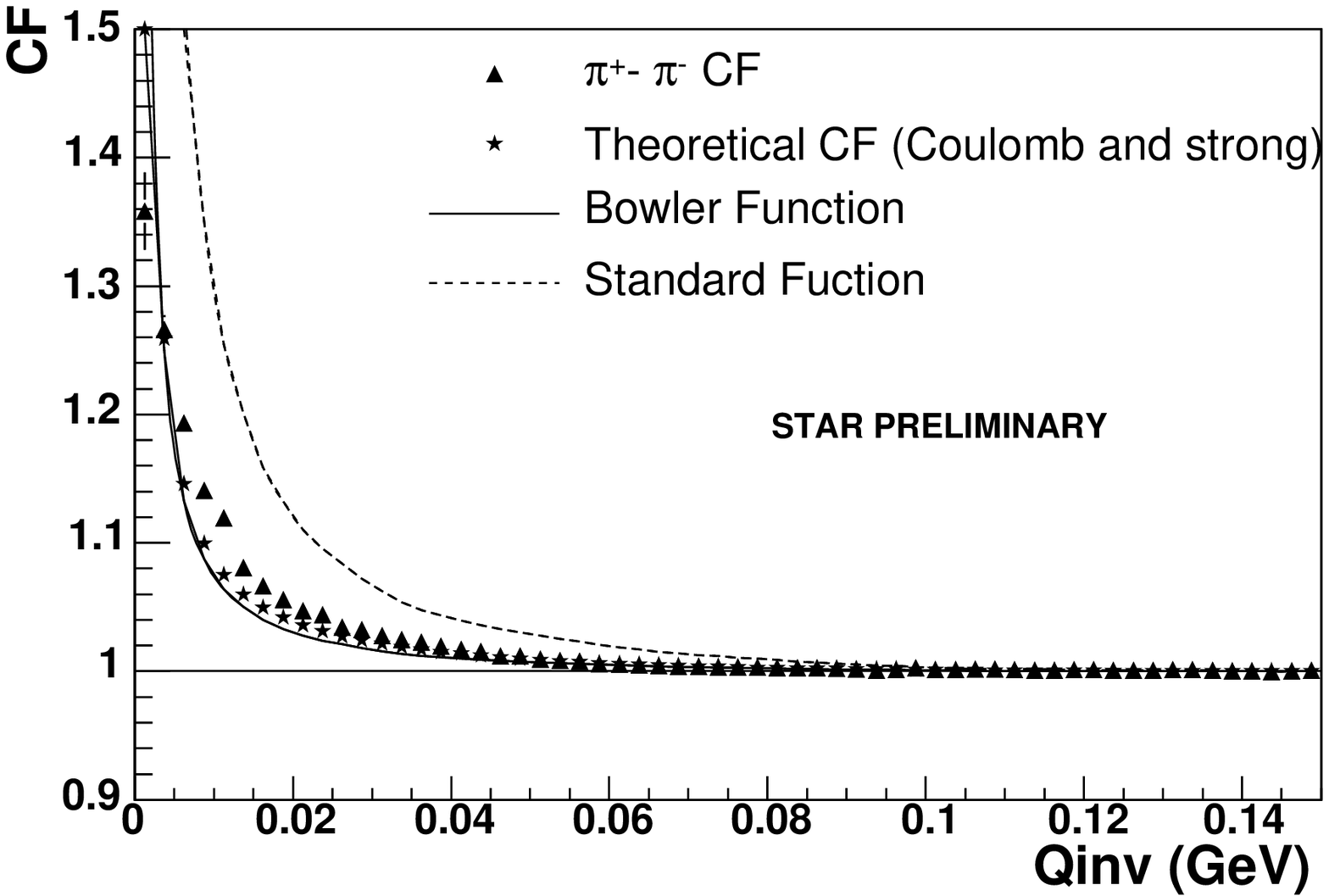}
\caption{Experimental (triangles) and theoretical \cite{LEDNI82}
(stars) 1D $\pi^{+} \pi^{-}$ correlation functions compared with
\textit{standard} (discontinuous line) and
\textit{Bowler-Sinyukov} (continuous line) functions.}
\label{QinvCF}
\end{minipage}\hfill
\begin{minipage}[t]{0.45\textwidth}
\includegraphics[width=\textwidth,height=1.00\textwidth]{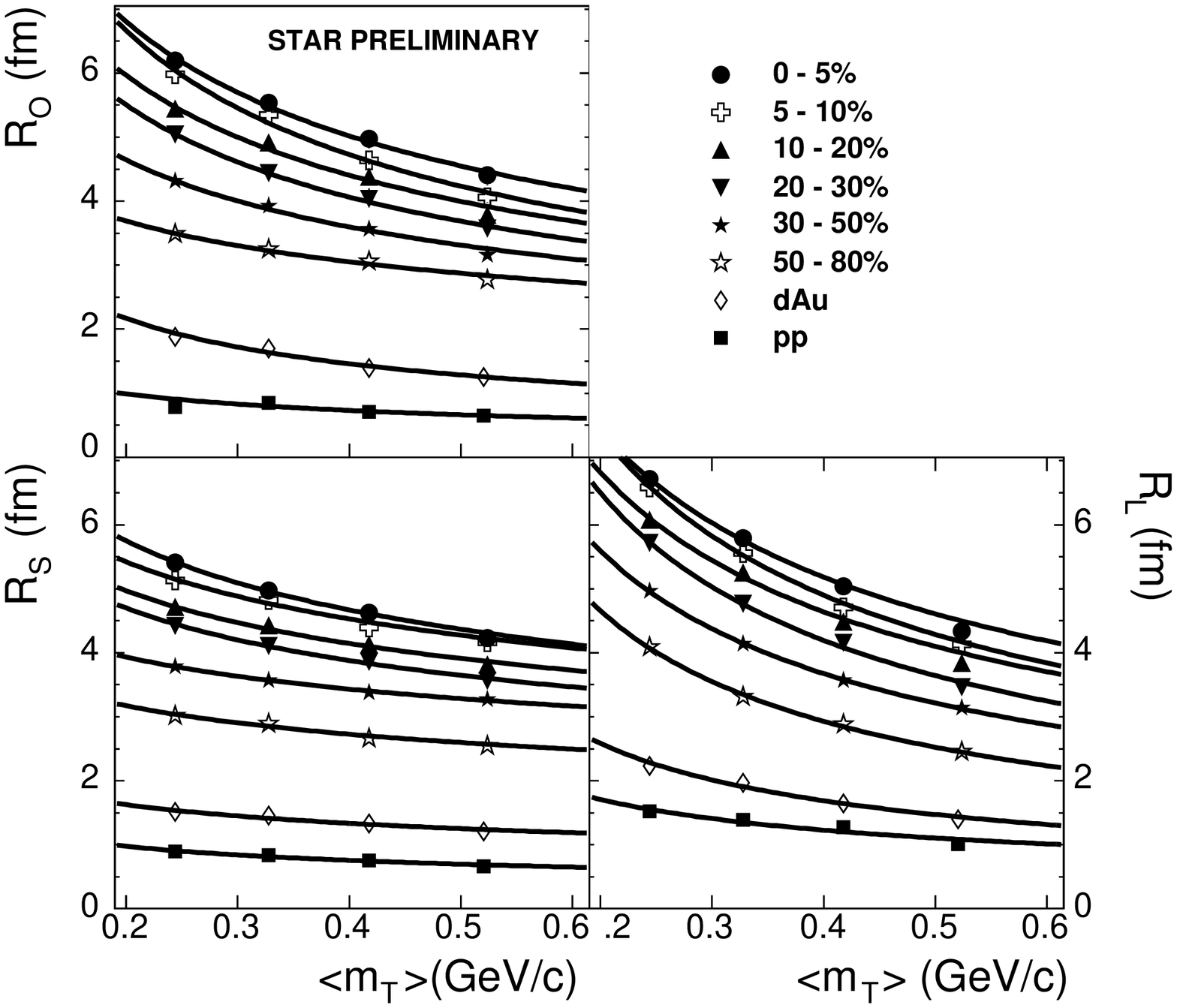}
\caption{HBT radii for 6 different centrality from Au+Au
collisions, and from p+p and d+Au collisions. The lines indicate
power-law fits to each parameter and centrality.}\label{HBTradii}
\end{minipage}\hfill
\end{figure}

In Figure \ref{QinvCF}, the measured $\pi^+\pi^-$ correlation
function is compared to several calculations.  Lines indicate the
\textit{standard} ($K_{coul}(Q_{inv})$) and
\textit{Bowler-Sinyukov} ($1+(\lambda-1)K_{coul}(Q_{inv})$)
Coulomb functions; in the latter, $\lambda$ was extracted from the
fit to the 3D like-sign correlation function.  Clearly, the
\textit{Bowler-Sinyukov} function better reproduces the data.
Further improvement is observed when the strong interaction
(negligible for like-sign pion correlations) is included
\cite{LEDNI82} into the $\pi^+\pi^-$ final state interactions.

When we use this procedure in our 3D analysis we observe an
increase in $R_{o}$ of 10-15$\%$. The values of $R_s$ and $R_l$ do
not depend significantly on the Coulomb procedure. Consequently
the increase in $R_{o}/R_{s}$ is not enough to solve the HBT
puzzle.

\section{HBT parameters versus centrality and transverse momentum}

\begin{figure}
\begin{minipage}[t]{0.48\textwidth}
\includegraphics[width=\textwidth,height=0.65\textwidth]{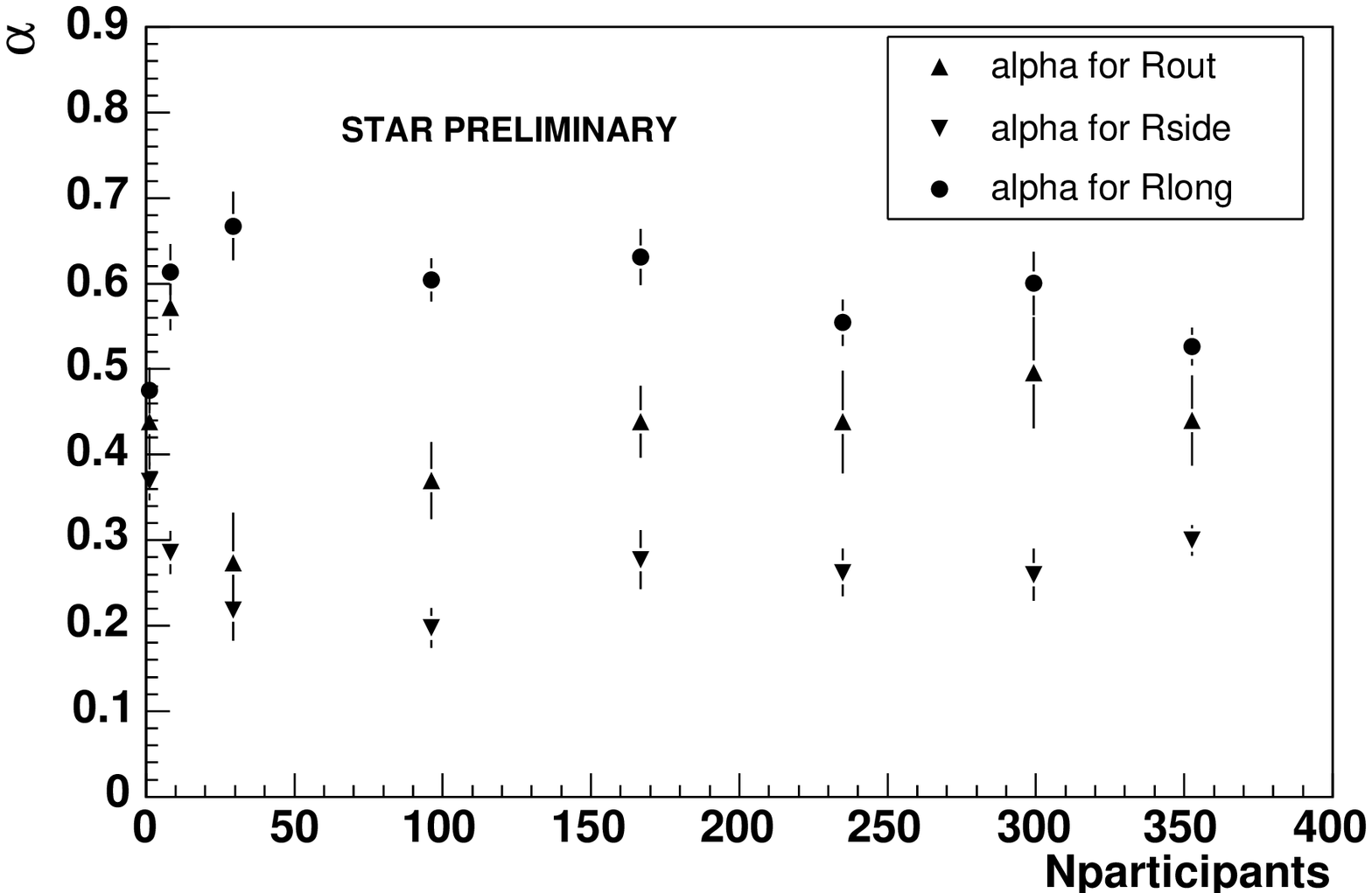}
\caption{Extracted $\alpha$ parameter from the power-law fits to
the HBT radii (lines in Figure \ref{HBTradii}) for p+p, d+Au and 6
different centralities in Au+Au collisions} \label{alpha}
\end{minipage}\hfill
\begin{minipage}[t]{0.48\textwidth}
\includegraphics[width=\textwidth,height=0.65\textwidth]{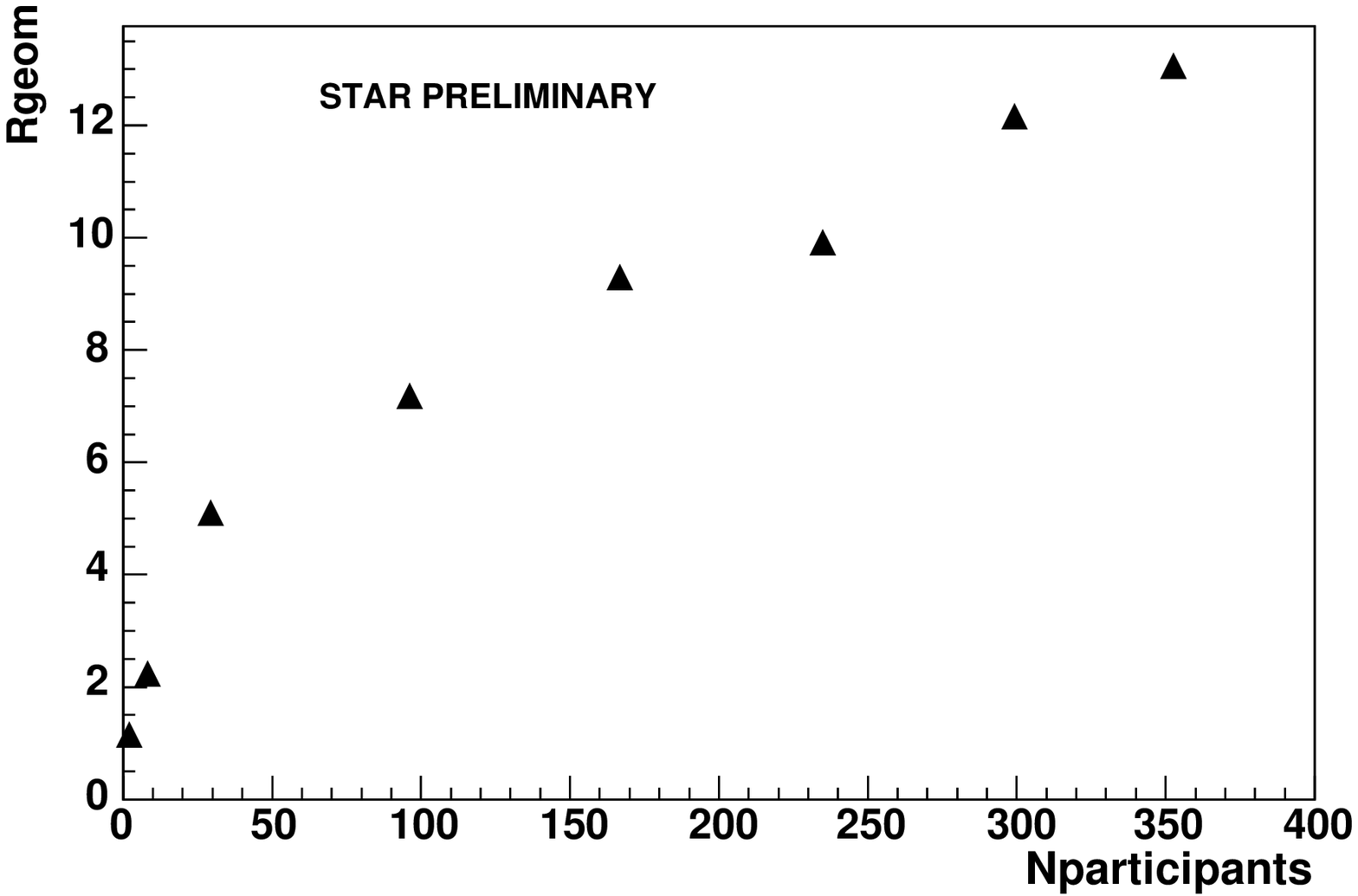}
\caption{Extracted $R_{geom}$ radius for a transverse gaussian
density profile \cite{WIEDE96} for p+p, d+Au and 6 different
centralities in Au+Au collisions}\label{Rgeom}
\end{minipage}\hfill
\end{figure}

Figure \ref{HBTradii} shows the $m_{T}$ dependence of the source
parameters for pions at six centrality bins from Au+Au collisions
as well as from p+p and d+Au collisions at same beam energy. The
three radii increase with increasing centrality and $R_{l}$ varies
similar to $R_{o}$ and $R_{s}$; for $R_{o}$ and $R_{s}$ this
increase may be attributed to the geometrical overlap of the two
nuclei. The extracted radii rapidly decrease as a function of
$m_{T}$, which is an indication of transverse flow \cite{TOMAS00}.
In order to extract information about the $m_{T}$ dependence on
centrality, we fit the $m_{T}$ dependence of each radius and each
centrality to a power-law function: $R_{i}(m_{T})~=~R_{i0}\cdot
m_{T}^{-\alpha}$ (solid lines in Figure \ref{HBTradii}). Figure
\ref{alpha} shows the dependence of $\alpha$ on the number of
participants; for Au+Au, $\alpha$ is constant for $R_{l}$ as a
function of number of participants and decreases with the number
of participants for $R_{o}$ and $R_{s}$ for the most peripheral
bins indicating a decrease of transverse flow for these
collisions. $R_{o}/R_{s} \sim$ 1 which indicates a short emission
duration in a blast wave fit \cite{RETIE03}.

Assuming boost-invariant longitudinal flow we can extract an
evolution time-scale by using a simple fit \cite{RETIE03}: $R_{l}
= <t_{fo}>
\sqrt{\frac{T}{m_{T}}\frac{K_{2}(m_{T}/T)}{K_{1}(m_{T}/T)}}$ where
T is the freeze-out temperature and $K_{1}$ and $K_{2}$ are the
modified Bessel functions of order 1 and 2. For T extracted from
fits to pion, kaon, and proton transverse momentum spectra (90 MeV
for most central and 120 MeV for most peripheral collisions)
\cite{ADAMS03_2} we get $<t_{fo}> \approx$ 9 fm/c for central
events and $<t_{fo}> \approx$ 6 fm/c for peripheral events. Hence,
the evolution time, in addition to the emission duration, is quite
short.

For a transverse expanding, longitudinally boost-invariant source,
and assuming a Gaussian transverse density profile, we can extract
information about its radius, $R_{geom}$, by fitting the $m_{T}$
dependence of $R_{s}$ to: $R_{s}(m_{T}) =
\sqrt{\frac{R_{geom}^{2}}{1 + \eta_{f}^{2}(\frac{1}{2} +
\frac{m_{T}}{T})}}$ where T is again the freeze-out temperature
and $\eta_{f}$ is the surface transverse rapidity \cite{WIEDE96}.
For T and $\eta_{f}$ consistent with spectra we see an increase on
this radius from $\sim$5 fm for the most peripheral case to
$\sim$13 fm for the most central one as shown in Figure
\ref{Rgeom}. We also observe a smooth transition from p+p
($N_{participants}$ = 2) and d+Au ($N_{participants}$ = 8.3) to
Au+Au collisions.

\section{Conclusion}

We have presented identical pion interferometry results for Au+Au
collisions at {\sqrtsNN}=200 GeV. With respect to multiplicity and
$m_{T}$ dependencies, pion HBT radii are very similar to results
reported at {\sqrtsNN}=130 GeV. HBT radii and geometrical radius
increase with increasing centrality. HBT radii decrease with
$m_{T}$ and we observe a stronger flow for the most central
collisions. Our results indicate that both the evolution timescale
(as measured by the $m_{T}$ dependence of $R_{l}$) and the
emission duration (probed by comparing $R_{o}$ to $R_{s}$) are
surprisingly fast. The Bowler-Sinyukov Coulomb procedure does not
solve the ``HBT puzzle" although increases the ratio $R_{o}/R_{s}$
by 10-15\%.

\end{document}